# A Novel Compaction Approach for SBST Test Programs

Juan-David Guerrero-Balaguera, Josie E. Rodriguez Condia, Matteo Sonza Reorda
Department of Control and Computer Engineering
Politecnico di Torino, Torino, Italy
{juan.guerrero, josie.rodriguez, matteo.sonzareorda}@polito.it

*Abstract* —In-field test of processor-based devices is a must when considering safety-critical systems (e.g., in robotics, aerospace, and automotive applications). During in-field testing, different solutions can be adopted, depending on the specific constraints of each scenario. In the last years, Self-Test Libraries (STLs) developed by IP or semiconductor companies became widely adopted. Given the strict constraints of in-field test, the size and time duration of a STL is a crucial parameter. This work introduces a novel approach to compress functional test programs belonging to an STL. The proposed approach is based on analyzing (via logic simulation) the interaction between the micro-architectural operation performed by each instruction and its capacity to propagate fault effects on any observable output, reducing the required fault simulations to only one. The proposed compaction strategy was validated by resorting to a RISC-V processor and several test programs stemming from diverse generation strategies. Results showed that the proposed compaction approach can reduce the length of test programs by up to 93.9% and their duration by up to 95%, with minimal effect on fault coverage.

## I. INTRODUCTION

Currently, innovative technologies and electronic products can provide several benefits, simplifying how we interact with society and increasing our life quality. The new products are often processor-based systems that include one or more processors to perform a given task [1].

Safety-critical products require effective in-field testing strategies, and one possible solution corresponds to the Software-Based Self-Test (SBST) strategy, which is based on the execution of special test programs (TPs) to detect faults. The SBST technique is commonly employed as a functional at-speed test mechanism to detect faults resorting to a test run at the maximum operating clock frequency. This technique is suitable to be applied for in-field testing during the operative life of a device and allows in particular the periodic testing of most internal modules. Several manufacturers currently use SBST, including MicroChip, Infineon, Cypress, ARM, Renesas, NXP and STMicroelectronics, thus providing in-field test capabilities for their products when used in several domains (such as industrial, medical, aerospace, and automotive) [2] [3] [4] [5].

A STL may include several TPs developed with different approaches and programming styles to achieve high fault coverage (FC). Each TP generally requires a limited amount of time to be executed. However, the application's constraints may limit the available time to perform periodic testing. In this scenario, fast TPs are desired. However, SBST routines have different characteristics, being generated with different approaches (e.g., custom, pseudorandom, deterministic, and ATPG-based), so the compaction of a TP can be a challenging task.

Several works have proposed different algorithms to reduce the duration of a TP while maintaining the same FC [6]. Some works proposed splitting the TPs into sub-routines and remove individual instructions [7]. Similarly, other works exploit reordering techniques among different pieces of a TP. This reordering technique can use a few parts of the original TP to keep the FC and reduce the length [8].

Nevertheless, a high computational effort is needed to analyze and compact a TP in most of the proposed works. Moreover, none of the proposed techniques considers the micro-architectural behavior of each instruction and its local fault propagation capabilities. Instead, they use high-cost approaches to create new TPs as compacted candidates versions of the original TP. For each new TP, a fault simulation is performed to assess the FC of the created TP and its length. Therefore, the cost required for the compaction of a TP is exceptionally high in terms of compaction time and computational resources required.

In this work, we propose a novel compaction approach, which is based on the observation that a significant part of the SBST TPs is composed of sequences of instructions having a structure composed of 3 parts:

1. One devoted to writing some specific values in suitable registers;
2. One using these values to stimulate some target module;
3. One part to make the results produced by the target module visible on some observable point.

Many test procedures addressing large combinational units (e.g., adders, multipliers, decoding units) have this structure. They are generated by transforming test patterns generated by a combinational ATPG into instructions or resorting to evolutionary or pseudorandom approaches.

More in detail, the novel compaction approach for SBST TPs addresses, in particular, those TPs having the above organization. This method analyzes the behavior of a TP at the micro-architectural level during its execution, combined with the fault detection ability of each test pattern generated by each instruction on the inputs of a combinational block and able to propagate any existing fault effect to some visible point.

The compaction method combines different levels of abstraction. First, a logic simulation using the RTL model of a processor is used to gather detailed tracing information about the executed TP for every clock cycle. Then, a fault simulation is performed using the gate-level version of the circuit. This fault simulation also records the number of faults detected at each clock cycle. The results obtained in these simulations are used to identify the instructions to be removed from the TP when they do not detect faults. In the end, this strategy reduces the required time for producing a compacted TP and has a minimum impact on the final FC.

The proposed method is highly effective in performing the compaction of TPs with continuous and regular structures. Additionally, the computational cost of this compaction approach requires only ONE fault simulation, drastically reducing the time required for the compaction of large TPs.

For the purpose of this work, we used the RI5CY processor and several SBST TPs to validate the proposed compacting approach and quantitatively evaluate its effectiveness. Results show that the compaction strategy can reduce the length of a TP by up to 93.9% and its duration up to 95.07% while causing a minimal reduction in the fault detection capabilities (<0.4 %).

The paper is organized as follows: Section II introduces the essential background and related works in the field. Section III describes the proposed approach. Section IV presents the study cases used to evaluate the proposed SBST compaction approach. Section V reports some experimental results, and Section VI finally draws some conclusions.

## II. BACKGROUND AND RELATED WORKS

### A. Software-Based Self-Test

SBST [9] is a flexible and noninvasive strategy that allows fault detection in the internal modules of a processor-based system. SBST can be applied at the end of the production phase and is also widely employed for in-field test. This strategy is based on executing specially crafted TPs using selected instructions at maximum operational clock speed. Each TP is built using the available Instruction-Set Architecture (ISA) of the target processor. Each instruction applies one or more test patterns to a target module. These instructions compose several routines aiming at exciting, propagating and detecting faults. Fault detection is commonly performed using signatures out of the values produced on any observation point or output of the processor. A comprehensive overview of the issues (and possible solutions) to be faced when generating STLs in an industrial environment can be found in [10].

### B. Related works

In the literature, several techniques have been proposed to reducing the size and duration of a TP without affecting the FC features. Most of these techniques are based on complex evolutionary algorithms, instruction classification and removal, or a combination of both.

Authors in [6] proposed a compaction method based on an evolutionary approach. This method transforms a TP into several small segments (or *spores*). Each spore program detects a portion of faults from the original program and, once combined, detects the same number of faults as the original program. Then, an evolutionary procedure is performed to get a subset of test patterns but keeping the same FC. A similar compaction evolutionary technique was introduced in [7]. In this work, the authors exploit the evolutionary technique to remove redundant instructions without affecting the achieved FC.

In [11], the authors presented a static compaction technique for TPs. In this approach, a set of TPs are analyzed and reduced to a sub-set of the original ones which maximizes the performance and also maintains the same global FC. Authors in [12] [13] presented two compacting methods based on the removal of instructions from TPs. The first method is called *compaction by instruction removal* (A0) and removes one instruction at a time from the original program. Then, a fault simulation is performed to observe the effect in terms of FC. When the FC is reduced, the removed instruction is reinserted in the program. Otherwise, this instruction is deleted permanently. This procedure is repeated for all available instructions in a TP. The second method is called *Restoration-based Algorithm* (*A1xx*) and initially splits the TP into small blocks. Then, one block of instructions is removed and each instruction is individually added in the main program to reach the original FC. In such a case, if one instruction inside the block does not increase the fault detection capabilities, it is removed permanently.

Finally, the authors in [14] described a compaction mechanism for TPs. The method is composed of two stages. The first stage identifies redundant instructions using a dependency data graph technique. The second stage reuses the *A1xx* approach, described above, and exploits optimization approaches to reduce the computational cost when executing the compaction algorithm.

Most previous works achieve TP compaction resorting to complex optimization algorithms and exploit a high-level abstraction, targeting the reduction of any TP without correlating the microarchitecture execution and the type of instructions employed. Moreover, these methods usually require a high amount of fault simulations, proportional to the length of the original TP, which is also proportional with the compaction costs.

In the present work, we face those compaction costs issues by extracting microarchitectural information during a preliminary logic simulation and determining the relation between the fault detection capabilities of each instruction in a TP, so identifying the main candidates for compaction and also reducing the compaction time needed. Moreover, the compacting approach exploits the microarchitecture-instruction-fault relation to reduce TP size and duration with minimal impact on the final FC.

## III. PROPOSED COMPACTION APPROACH

The proposed approach assumes that a Self-Test Library for a given processor is available and can be divided into TPs (TPs). Each TP is composed of a given number of instructions, lasting for a certain amount of clock cycles, and achieving a certain FC with respect to a given fault model. For the purpose of this paper we focused on the stuck-at fault model, although the method can be easily extended to different fault models. Moreover, these TPs can be split into sets of instructions (Basic Blocks or BBs), each corresponding to a consecutive group of instructions, which are always executed in sequence (no branches or jumps in/out the BB) [15].

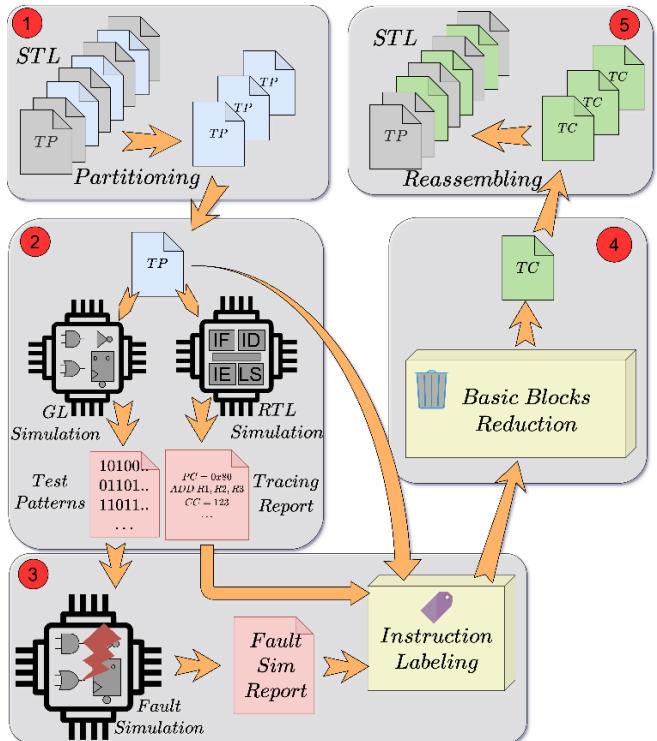

**Fig. 1.** A general scheme of the proposed compaction approach for functional TPs.

The proposed compaction procedure only uses one RTL logic simulation to collect tracing information about the original TP targeting a given device (i.e., the CPU). This tracing information contains for each clock cycle the decoded instruction and the value of the program counter. Then, one fault simulation is performed on the TP and one report generated. This report and the trace information are used to identify the instructions to be removed. More in detail, the compaction approach is divided into five stages: *i)* Test program partitioning, *ii)* Logic tracing, *iii)* Fault detection analysis and instruction labeling, *iv)* Program reduction, and v) Test program reassembling. Figure 1 shows the steps and the main flow of the proposed compaction approach.

Initially, in the test program partitioning stage, the TPs (of an STL) are analyzed and classified into two groups: *1)* those that can be compacted, and *2)* those which are not suitable. The first group is composed of TPs, including at least one unconditional BBs. Those unconditional BBs are executed once and does not depend on any conditional assessment to be executed. Therefore this method considers those regions in a TP (composed of unconditional BBs) as *admissible* regions [8], other structures are not candidates for compaction and remains unaffected. This partitioning process can be easily automated. Moreover, the other stages only analyze the admissible regions of each TP.

The logic tracing stage aims at collecting information to identify the relationship between each instruction in each BB and its effects in terms of fault detection. In this stage, each TP is analyzed using one RTL logic simulation and one Gate-level (GL) logic simulation.

---
**Instruction labeling algorithm**

**Input**: test program TP composed of I instructions, Tracing clock cycle report CC, Tracing program counter report PC, Tracing decoded instruction report DI, Fault sim test patterns report FSR
**Output**: Labeled test program TPL
**for** each clock cycle *k* in CC **do**
| **if** FSR(*k*) detects faults **then**
| | **if** DI(k) matches TP(PC(*k*)) **then**
| | | TPL{k}:= ('*essential*', TP(PC(k)))
| **else**
| | TPL{k}:= ('*not essential*', TP(PC(k)))
**loop**

**Fig. 2.** Instruction labeling algorithm.

---

On the one hand, the RTL logic simulation provides the main execution information about the TP with respect to the target device per clock cycle. This simulation produces one tracing report that captures the detailed information of the HW-SW interface. The tracing report contains the execution sequence for the analyzed TP in terms of the program counter value, the decoded instruction, and the clock cycle. One hardware monitor inserted in the RTL model of the device generates the report and captures the tracing information in the execution pipeline stage. The results of the executed instruction can be observed in the memory bus system or any other observation point. Therefore, any malfunction caused by a fault can be identified at this point without trace each pipeline stage.

On the other hand, the GL logic simulation runs each TP and generates the input sequence of logic values per clock cycle (*Test Pattern*). These test patterns are extracted as I/O switching activity for the target device and are employed by the fault simulation in the next stage.

The fault detection analysis stage is divided into two steps. In the first step, a fault simulation is performed on the Gate-level description. This fault simulation employs, as inputs, the test patterns generated by the GL logic simulation in the previous stage. Moreover, the fault observation point is restricted to the memory bus system. During the fault simulation, one fault is detected when there is a difference between the fault-free system and the current faulty system at any clock cycle. The fault simulation generates a report containing the list of all test patterns and the number of faults activated and detected by each test pattern. It is worth noting that the proposed compaction method only requires ONE fault simulation for each TP under analysis.

The second step correlates the fault simulation report and the collected trace information from the RTL logic simulation. The information from the fault simulation report is used to identify the test patterns corresponding to each instruction. At the same time, this matching procedure identifies the fault detection capabilities of each instruction of the analyzed TP. The instruction matching takes the executed instruction, which is registered in the tracing report on a clock cycle (*k*), and the number of faults detected when a test pattern is simulated in the same clock cycle *k*. if there are detected faults on the *k* clock cycle, the instruction traced is located in the source code and is labeled as "*essential*". Otherwise, the instruction is labeled as "*not essential*" and is a candidate instruction to be removed from the program. Figure 2 describes the labeling procedure algorithm.

| Reduction Algorithm |
|---|
| **Input**: Labeled Test program TL with I instructions divided into m consecutive BBs (BB1, BB2, BB3, . . ., BBm); |
| **Output**: Compacted test program TC |
| **for** each basic bloc BBx in TL **do** |
|   \|   **for** each instruction in BBx **do** |
|   \|   \|   **if** label of instruction is *essential* **then** |
|   \|   \|   \|   TC{:}:=BBx |
|   \|   \|   \|   continue to next basic block |
|   \|   **loop** |
| **Loop** |

**Fig. 3.** Reduction algorithm applied to test programs used to remove basic blocks.

The fourth stage (Program reduction) analyzes and reduces the labeled TPs. More in detail, this method examines, instruction by instruction, each BB in a given TP. When at least one instruction inside the BB is marked as "*essential*", this BB cannot be removed from the TP. Then, those BBs in which all instructions are labeled as "*not essential*" can be removed from the TP, only. Figure 3 describes the proposed reduction algorithm used in the compaction approach.

The reassembling stage replaces the previous version of the TP using the new one, obtained from the compaction procedure. Finally, a fault simulation of the compacted TP or STL is performed to evaluate the quality of the new TP.

## IV. STUDY CASES

We employed the RI5CY processor core of the Parallel Ultra Low Power (PULP) platform as the target device to evaluate and validate the proposed compaction approach. It is worth noting that the proposed compaction approach can be adapted to any other processor.

The RI5CY is a 4-stage in-order RISC-V processor core and was synthesized using the 45nm Nangate OpenCell library [16]. Table I reports the list of modules and the number of stuck-at faults in each module on the RI5CY processor.

We selected four STLs and applied our method on them. Each STL includes several TPs written in assembly language to test the RI5CY core. It is worth noting that skilled test engineers designed all functional TPs using different approaches targeting the main component of the CPU (Execute unit, Registers File, Instruction decode, Instruction fetch, and Load Store Unit).

The STLs include TPs (TEx) with admissible regions specially designed to test the execution unit of the CPU. For the purpose of this work, we extracted these TP (TEx) to evaluate the compaction effect on both individual TPs and complete STLs. Table II reports the main features of the selected STLs and TPs. This table includes the size, the duration in clock cycles (cc), the FC, and the percent size of the admissible region of each TP, according to the constraints explained in Section III. In Table II, the admissible region per TP varies as the ability to apply the compaction approach. As observed, the minimum percent size of the admissible region is 41.75%.

TABLE I. NUMBER OF FAULTS PER MODULE IN THE RI5CY PROCESSOR

| Module in the CPU | Number of faults |
|---|---|
| Instruction Fetch | 24,148 |
| Instruction Decode | 50,340 |
| Execute | 63,878 |
| LSU | 4,442 |
| CS Registers | 6,958 |
| Frontend | 11,270 |
| Full CPU | 161,036 |

TABLE II. MAIN FEATURES OF THE CONSIDERED TEST PROGRAMS

| Benchmark | Size (instructions) | Admissible region (%) | Duration (cc) | FC (%) |
|---|---|---|---|---|
| STL1 | 13,845 | 41.75% | 126,706 | 82.97 |
| STL2 | 64,390 | 97.04% | 374,138 | 85.04 |
| STL3 | 91,623 | 82.87% | 977,012 | 85.04 |
| STL4 | 218,467 | 98.16% | 601,966 | 86.59 |
| TE1 | 5,780 | 100% | 8,589 | 88.14 |
| TE2 | 33,034 | 100% | 79,152 | 91.68 |
| TE3 | 60,594 | 100% | 84,580 | 94.93 |
| TE4 | 206,306 | 100% | 439,954 | 95.84 |

STL1 is composed of three TPs, each TP targeting individual modules in the CPU. A March algorithm is used to test the register file and the control and status registers of the CPU [17].

The execution units of the CPU are tested by STL1, using TE1, which comprises instructions targeting the individual test of internal units (ALU, DotP Unit, Multiplier unit, and Divider Unit). Each instruction in TE1 was generated using an ATPG tool: each ATPG-generated test pattern was turned into one or a sequence of instructions for the TE1.

STL2 is composed of several test routines combined in seven independent TPs. The TPs for the register file and the execute unit were developed using a similar procedure as in STL1. One March algorithm targets the test of the register files and ATPG generated test patterns (TE2) target the execution unit. Some additional TPs are included using nested loop-based algorithms to take advantage of the CPU's hardware loops feature. Those loops use multiply and divide instructions and contribute to increase the FC in the execute unit.

STL3 comprises six independent TPs. It uses a similar March algorithm to test the register file in the CPU, as STL1. The execution unit is tested using pseudorandom instructions. TE4, the TP for the execution units, was designed using a custom pseudorandom generator tool. More in detail, TE4 is built using a BB with a fixed structure and size. Each BB contains three parts: *i)* register initialization using random data, *ii)* instruction selection (selecting one random instruction taken from the available ISA of the CPU). In this case, branch and control-flow instructions are discarded, and *iii)* fault propagation, using one store instruction to propagate any fault effect to one of the available observation points. Each instruction's source and destination registers are selected randomly, and each register can only be used once.

Additionally, one small TP based on hardware loop core increases the fault detection in the multiplier and the divider, similar to STL1 and STL2.

STL4 incorporates two independent TPs: one for the execution unit (TE4) and one for all the other modules. TE4 follows a similar construction approach as TE3, but each BB has a random length (sets of 2 to 10 instructions). In STL4, the other modules are tested indirectly or adding special routines using loop-based algorithms and employing a special initial operand data as seed. In the case of STL2 and STL3, a pseudorandom approach is also employed. For STL1 and STL4 several manual parts were also added to complement the test of specialized modules, such as the Load-Store unit.

## V. EXPERIMENTAL RESULTS

The proposed compaction approach was implemented as a tool in Python language. This tool interacts with one commercial logic simulator and one commercial fault simulator framework to analyze and compact a given STL or TP. The experimental compaction results were obtained on a workstation with two AMD EPYC 7301 16-core processors running at 2.2GHz and equipped with 128 GB of RAM memory.

Initially, the individual TPs (TEx) were evaluated. Table III reports the main results of the compaction approach applied to each TP, considering only faults in the execute unit during the analysis for compaction. According to results, the proposed compaction approach can greatly reduce both the size (up to 93.9%) and the duration (up to 95.08%) for the evaluated TPs (TEx).

In principle, the observed compaction effects are directly related to the description style of the TPs and the capacity of each instruction to propagate any possible fault effect to one of the available observation points. In fact, the style of description defines the granularity of the BBs (few to hundreds of instructions). A deep analysis of the TPs revealed that all of them (TEx) use a few instructions per BB without control-flow instructions (the BBs in TE1 and TE2 are in the range of 6 to 8 instructions. TE3 have a size from 13 to 60, and TE4 has BBs with 3 to 6 instructions), so allowing a fine grain compaction by evaluating and possibly removing each BB.

TE1 has the shortest size among the analyzed TPs. Interestingly, the compaction was able to remove 1,916 ineffective instructions. Furthermore, the size of TE2, TE3, and TE4 was reduced by 82%, 91%, and 93%, respectively. Although TE1 and TE2 were created resorting to an ATPG, the compaction technique demonstrates good capacity in removing a significant amount of not essential instructions in the test.

Deep analysis of the long TPs (TE2, TE3 and TE4) revealed that the initial set of BBs mainly applies most test patterns (instructions), so the compaction of those BBs is not feasible. In contrast, BBs in the middle and at the end of the TP are redundant and contain less effective test patterns, allowing compaction and size reduction, as observed in Table III. Finally, the compaction also contributes to reducing the memory footprint of each TP by an identical percentage.

According to the results, there is a proportional relation between the percentage of reduced size and compacted duration of each TP. In fact, the high percentage of fine grain BBs in all TPs (100.0%) allow the compaction with similar percentage of reduction for both (size and duration). Our method can reduce up to 95.05% of the duration for the longest analyzed TP TE4. As explained above, several instructions produce redundant test patterns, which are weak in detecting faults from the execution unit. Then, those "redundant and weak" instructions are removed with minimal effect on the FC (see the differential (Diff) effects in FC in Table III).

TABLE III. THE COMPACTION RESULTS IN THE TEST PROGRAMS FOR THE EXECUTE UNIT

| Test Program | Compaction | | | | | |
|---|---|---|---|---|---|---|
| | Size | | Duration | | Diff FC | Compaction |
| | instr | (%) | (cc) | (%) | (%) | time (min) |
| TE1 | 3,864 | 33.15 | 5,860 | 31.77 | -0.07 | 5.32 |
| TE2 | 5,806 | 82.42 | 10,915 | 86.21 | -0.40 | 7.05 |
| TE3 | 4,999 | 91.75 | 7,299 | 91.37 | -0.11 | 10.52 |
| TE4 | 12,581 | 93.90 | 21,660 | 95.08 | -0.06 | 15.35 |

TABLE IV. THE COMPACTION PROCESS RESULTS OF EACH STL IN THE ADMISSIBLE REGION

| Benchmark | Compaction of the admissible region of STL | | | | | |
|---|---|---|---|---|---|---|
| | Size | | Duration | | Diff FC | Compaction |
| | instr | (%) | (cc) | (%) | (%) | Time (hours) |
| STL1 | 3,864 | 33.15 | 5,860 | 31.77 | -0.07 | 0.28 |
| STL2 | 33,706 | 46.06 | 182,456 | 32.95 | +0.08 | 7.16 |
| STL3 | 18,939 | 75.06 | 76,745 | 82.82 | -0.11 | 8.04 |
| STL4 | 16,263 | 92.42 | 25,436 | 84.95 | -0.29 | 4.18 |

In all the analyzed TPs, both parameters (size and durations) were reduced. Meanwhile, the FC is the most relevant parameter, and it is expected to remain constant. The results show that the FC of each TP is reduced by a small percentage (from 0.01% to 0.4%). This minimal reduction in the FC is caused by the data dependence between consecutive BBs and weak test patterns. This behavior appears when a BB uses the results of a weak and removed BB as inputs, so the missing BB does not perform the required operations to generate the inputs for the "essential" BB. In the end, the fault detection decreases in those BBs that depend on the operation of other BBs. In the analyzed TPs, the previous scenario was less frequent than 1% of all BBs.

Then, the compaction approach was applied to complete STLs. In these cases, the admissible region of all TPs of a given STL (two TPs for STL1, thirteen for STL2 and STL3, and five for STL4) were analyzed and compacted.

The compaction method was applied independently to each TP. The results reported in Table IV show the minimum compacting impact on FC (<0.29%) for complete sets of TPs. Interestingly, in STL2, the FC was increased by 0.08%; this happens when one BB uses the propagated data of one removed BB. This remotion creates a favorable test pattern that detects additional faults. Additionally, in the admissible region of the STLs, the compaction approach reduced the size in the range from 33.15% to 92.42%. Similarly, the duration was reduced in the range of 31.77% to 84.95%.

According to the results in the admissible region, the percentage of the reduction in size and duration is slightly lower with respect to the reduction of TEx, only. This difference is caused by a bigger granularity in the BBs. In case of at least one essential instruction inside a BB, this BB must be preserved in the TP or STL, so complicating the complete removal of BBs in the STLs, especially for STL2 with BBs in sizes between 60 and 135 instructions.

TABLE V. THE COMPACTION RESULTS IMPACT IN THE ENTIRE STL

| Benchmark | Compaction STLs | | | | | |
| --- | --- | --- | --- | --- | --- | --- |
| | Size | | Duration | | Diff FC | Compaction |
| | instr | (%) | (cc) | (%) | (%) | Time (hours) |
| STL1 | 11,929 | 13.84 | 123,977 | 2.15 | −0.07 | 0.28 |
| STL2 | 37,513 | 41.74 | 287,967 | 23.03 | +0.08 | 7.16 |
| STL3 | 50,322 | 45.08 | 884,973 | 9.42 | −0.11 | 8.04 |
| STL4 | 24,314 | 88.87 | 178,985 | 70.27 | −0.29 | 4.18 |

Table V reports the main features of the compaction of the complete STLs. From results, STL4 has a good compaction rate in duration (70.27%) and size (88.87%). On the other hand, the size reduction achieved for STL1, STL2 and STL3 is moderate (in the range of 13% to 45%). Unfortunately, the duration reduction in STL1 and STL3 is low (<10%) and moderate for STL2 (23.03%). It is worth noting that the high percentage of the inadmissible region in STL1 and STL3 represents more than 50% of the total execution time for both STLs, which is the main cause of the low percentage of reduction of the analyzed STLs.

The required time by our compaction method is determined by the time of one fault simulation. The compaction of a TP targeting faults in the execution unit of the CPU takes from 5 to 12 minutes. In contrast, when an STL is analyzed for compaction and all faults in the CPU are considered, the required time depends on the number of TPs and the duration of each one of them. In our experiments, the maximum time required for the compaction of a complete STL reaches 8.04 hours.

It must be noted that each analyzed TP included thousands of instructions. Moreover, the target processor has more than a hundred thousand faults. For this features the proposed compaction method requires only one logic simulation and one fault simulation. In contrast, several works [6],[7],[10],[13],[14],[8] compacted STLs and TPs using techniques that require as many fault simulations as the number of instructions in a TP. In the end, for those methods, the required compaction time is proportional to the number of fault simulations, usually in the order of hundreds or thousands of fault simulations.

## VI. CONCLUSIONS

In this paper, we propose a novel approach to compact the size and duration of STLs and functional test programs described using the SBST strategy. The compaction approach addresses test programs with a regular structure of consecutive instructions or basic blocks. As observed in the results, the proposed method greatly reduces the required compaction time by exploiting only one fault simulation per test program. The proposed approach was validated by resorting to a state-of-art pipelined microprocessor.

The compaction method combines different levels of abstraction by combining one RTL logic simulation and one GL fault simulation to extract, trace and label the portions of a test program that can be removed. Then, after analyzing the essential instructions per basic block, those non-essential instructions are removed from a given test program.

Despite the fact that the compaction approach has some constraints, results showed that this method can reduce the size (by up to 93.9%) and the duration (by up to 95.08%) for the test programs targeting the execution unit, and up to 88.87% reduction in size and 70.27% reduction in duration for STLs developed to test the entire CPU. Finally, the compaction method has a negligible impact on the FC (<0.4%), while we also observed cases where the FC increased.

As future works, we plan to extend the compaction capabilities to more general programming styles. Moreover, we plan to explore the compaction of functional test programs for parallel processors and GPUs.